\documentclass[10pt]{article}

\usepackage{amsmath}

\usepackage{xcolor}

\begin{document}

\title{From Symmetry to Supersymmetry to Supergravity\thanks{Invited contribution to {\it Half a Century of Supergravity}, A. Ceresole and G. Dall'Agata (eds.), Cambridge University Press.}}

\author{ {Elena Castellani\thanks{Department of Humanities and Philosophy, University of Florence, via della Pergola 60, 50121,
Firenze, Italy. E-mail: elena.castellani@unifi.it}}}

\date{}

\maketitle

\section{Introduction}

The theoretical developments that led to supersymmetry -- first global and then local -- over a period of about six years (1970/71-1976) emerged from a confluence of physical insights and mathematical methods drawn from diverse, and sometimes independent, research directions. Despite these varied origins, a common thread united them all: the pursuit of a unity in physics, grounded in the central role of symmetry, where ``symmetry'' is understood in terms of group theory, representation theory, algebra, and differential geometry.\footnote{As Wess (2000, 67) nicely put it,  ``Everybody has his own concept of symmetry, also physicists, and you never know if we communicate about the same thing. Fortunately, mathematics with its strong capability to abstract has abstracted the concept of symmetries to the concept of groups.  When referring to symmetry I mean it in the framework of group theory, representation theory, algebra and differential geometry''.}

As is well known, the algebraic approach to physical symmetries -- by allowing the unification of various types of symmetries by means of combining their corresponding transformation groups into larger groups -- has provided the technical framework for the quest toward a unified theory of elementary particles and their interactions. From this perspective, the key question guiding the development of supersymmetry and supergravity was whether this strategy of enlarging the symmetry could achieve a {\it complete} unification: that is, a unified theory with symmetries relating not only particles with different internal quantum numbers (connected to charges of various types), as in the case of the Standard Model and Grand Unified Theories, but also particles of different spin (a property related to spacetime symmetry) -- namely, bosons and fermions.\footnote{As Freedman (1978, 535) phrased it, ``If supersymmetry or supergravity is the answer, then what is the question? There is a serious reply to this existential query, namely: is there a symmetry principle powerful
enough to raise hope for complete unification of the elementary particles and their interactions?''}

In other words, the ultimate goal was to seek a deeper connection between spacetime (external space) symmetries and inner space symmetries by enlarging the algebra in order to  
get a new symmetry that could change the statistics (Bose-Einstein vs Fermi-Dirac) of particles -- the generalized symmetry that would be called, in 1974, ``supersymmetry".
As was emphasized at the time, a more general symmetry relating bosons and fermions could also provide a way to address some fundamental questions 
about the divide between these two kinds of particles and the distinct roles each one plays in nature.\footnote{See for example the following recollection by Volkov (2000, p. 56): ``I would like to emphasize that questions that were also related to supersymmetry interested me already at the earliest stage of my activity. How do particles with different spin differ from one another? Why are there bosons, why are there fermions?  Questions that were also related to supersymmetry interested me [... ]  At the same time, the ideology of gauge
field started to penetrate actively into physics and I returned to the old questions. It amazed me that all these particles, the Goldstone particles and the gauge fields, were bosons, but the fermions somehow were not involved at all. Here, a kind of inequality appeared: why were some particles - bosons - selected, but others - fermions - not included in this group?''. Another example is the following reflection by Varadarajan (2010, 2): ``Supersymmetry was invented by the physicists to provide a unified way of understanding the behavior of the two basic constituents of the physical world, the fermions and the bosons.}

Up to the 70s, previous attempts at combining spacetime (external space) symmetries and inner space symmetries had failed, giving rise to a series of no-go theorems, culminating in the 1967 Coleman-Mandula theorem about all possible symmetries of the {\it S} Matrix. For a local, relativistic quantum field theory in
four-dimensional spacetime and on the basis of a given number of assumptions, this powerful theorem proved the impossibility of combining spacetime symmetries (Poincar\'e group) and internal symmetries (compact Lie groups) in any but a trivial way: in their authors' wording, it showed that the ``symmetry group of the {\it S}  Matrix is necessarily locally isomorphic to the direct product of an internal symmetry group and the Poincar\'e group'' (Coleman and Mandula, 1967, 159).  

In fact, the theorem was so precise that it suggested a unique way of avoiding its restrictions:\footnote{The fact that there is a unique symmetry circumventing the no-go theorem of Coleman and Mandula was shown by Haag, Lopuszahski and Sohnius (1975), proving that the supersymmetry algebra is the only graded Lie algebra of symmetries of the S-matrix consistent with relativistic quantum field theory.} it turned out that the only possibility of combining (non-trivially) external and internal symmetries was by generalizing the concept of symmetry, i.e. enlarging the algebra associated to the symmetry, thus relaxing one of the theorem's assumptions. In more detail, the idea for arriving at a symmetry between bosons and fermions was ``to generalize the notion of a Lie
algebra to include algebraic systems whose defining relations involve anticommutators as well as commutators" (Wess and Bagger, 1992, 4). 
These ``super-algebras'', already known in mathematics as {\it  graded Lie algebras}, represented extensions of ordinary Lie algebras in which a distinction was introduced between even elements (bosonic), obeying commutation relations, and odd elements (fermionic), obeying anticommutation relations among themselves and commutation relations with the even elements (e.g., Ferrara 1987, 3). 

Graded Lie algebras were not a new notion. They had been known since the mid-50s, first in the context
of the theory of deformation of algebraic structures (starting with Nijenhuis 1955), and later, in relation to mathematical questions of the second quantization, 
in the study of analogs of Lie algebras, differing from usual Lie algebras by properties of the commutator (Berezin and Katz, 1970).\footnote{A detailed review of graded Lie algebras in mathematics and physics until mid-70s is provided in Corwin, Ne'eman, and Sternber (1975).} The mathematical background for supersymmetry -- a new kind of algebraic structure, formulated in terms of commutators and anticommutators -- was thus already in place. However, the road to establishing a four-dimensional supersymmetric quantum field theory, first as global symmetry, then also as a local symmetry (supergravity), was anything but straightforward. The following two sections are devoted to illustrate the entanglement of ideas, methods and motivations informing the entire process:  ``from symmetry to supersymmetry" in Section 2,\footnote{``From Symmetry to Supersymmetry'' is the title of a 1999 lecture delivered by Julius Wess, reprinted in Kane and Shifman (2000).} ``from supersymmetry to supergravity'' in Section 3.

\section{From symmetry to supersymmetry}

According to the key figures in the developments leading to supersymmetry, the theory was ``discovered'' {\it independently} three times: 
two times in the former Soviet Union (in Moscow and Kharkov, respectively), the third time in the West. In the early 70s, that is well before the end of the Iron Curtain, communication between the two blocks was not always easy. However, it appears there was no real interaction also between the two soviet groups -- Y. A. Golfand and E. P. Likhtman in Moscow, and 
D. V. Volkov, V. P. Akulov, and V. A. Soroka in Kharkov.\footnote{See how Zumino remembers these early developments (2006, 199): ``It is remarkable
that Volkov and his collaborators didn't know about the work of Golfand and Likhtman, since all of them were writing papers in Russian in Soviet journals. Julius and I were totally unaware of the earlier work.''}

Although the overarching goal was unity in physics, the three pioneering lines of inquiry emerged from distinct motivations and starting points. 
To illustrate these developments, let us begin by outlining the main achievements that led to supersymmetry by listing the key authors and papers from 1971 to 1974.\footnote{Of course more papers appeared in the field than those listed here: the choice has been limited to those generally considered to be milestones in the progress towards supersymmetry in those years.} The details will then be examined in 2.1-2.2.

\medskip

$\bullet$ {\bf Soviet Union}:  Construction of supersymmetric field theories and studies of some of their properties. 

\smallskip

Physicists:  Y. A. Golfand, E. P. Likhtman, D, V. Volkov, V. P. Akulov,  V. A. Soroka.

\begin{itemize}

\item[$\bullet$] {\it Moscow}: 

Golfand and Likhtman, 1971 (``Extension of the Algebra of Poincar\'e Group
Generators and Violation of {\it P} Invariance'')

\smallskip

\item[$\bullet$]{\it Kharkov}: 

Volkov and Akulov. 1972 (``Possible Universal Neutrino Interaction''); 1973 (``Is the Neutrino a Goldstone particle?'')

\end{itemize}

\medskip

$\bullet$ {\bf West}:   From two-dimensional supergauges in the dual models of early string theory to four-dimensional supersymmetry.

\smallskip

Physicists: (1) P. Ramond, A. Neveu, J. Schwarz, J.-L. Gervais, B. Sakita; (2) J. Wess, B. Zumino.

\begin{itemize}

\item[$\bullet$]{\it Two-dimensional supergauges in dual models}: 

Ramond, 1971 (``Dual Theory for Free Fermions'')

Neveu and Schwarz, 1971 (``Factorizable dual model of pions'')

Gervais and Sakita, 1971 (``Field Theory Interpretation of Supergauges in Dual
Models'')

\item[$\bullet$]{\it Four-dimensional supersymmetry}:

Wess and Zumino, 1974a  (``Supergauge Transformations in Four Dimensions'')

\end{itemize}

\bigskip

\subsection{Soviet Union} 

\medskip

$\bullet$ {\it Moscow}.  The 1971 paper by Yuri Golfand and his graduate student Evgeny Likhtman at the Lebedev Physics Institute in Moscow is commonly acknowledged to be the first contribution where a four-dimensional supersymmetric field theory was obtained. 
In fact, Golfand had been working on extending the algebra of the Poincar\'e group since the late 1960s (Likhtman 2000, 35). As Golfand and Likhtman recollect,  ``Somewhat earlier one of the author (Yu. A. G.) considered spinorial extensions of Poincar\'e group wishing to come across new no-go theorems'' (Golfand and Likhtman (2000, 54). Thus, a motivation for these previous studies by Golfand was to deal with no-go theorems such as the Coleman-Mandula one.

After studying several kinds of extensions of the algebra by adding spinorial generators,  the purpose of the paper was  ``to find such a realization of the [enlarged] algebra ... in which the Hamiltonian operator describes the interaction of quantized fields'' (Golfand and Likhtman 1971, 323). They used Dirac spinors and found, in their realization, that the algebra seemed to be non invariant under space reflections. This was seen as a way to account for parity violation in the weak interactions -- whence the paper's title, ``Extension of the Algebra of Poincar\'e Group Generators and Violation of {\it P} Invariance''.\footnote{According to Marinov (2000, 166), for example, ``It is known that Golfand discussed the new symmetry with his colleagues in the late 1960's, trying to solve the puzzle of weak interactions, before the electroweak theory did the job.''}

After this first result -- in their own words, ``a model for the interaction of quantized fields with parity non conservation, invariant against the algebra'' (1971,  326) -- Golfand and Likhtman 
still worked on various aspects of supersymmetry. Likhtman, in particular, continued to work at new versions of interactions of the supermultiplets, and, according to his recollections, he was ``the first to observe that the number of the fermion and boson degrees of freedom coincided in every supermultiplet, while the infinite energy of vacuum oscillations is cancelled'' (Likhtman, 2000, 37). \\

\noindent $\bullet$ {\it Kharkov}. The second, independent discovery of supersymmetry is due to 
Dimitri Volkov and his graduate student Vladimir Akulov at the Kharkov Institute for Physics and Technology.  In their 1972 work,\footnote{The paper was sent both to the Russian {\it JETP Letters}, with the title ``Possible Universal Neutrino Interaction'', and to {\it Physics Letters} (where it appeared in 1973), with the title ``Is the Neutrino a Goldstone particle?''} 
by using the 1970 results of Beretzin and Katz on Lie groups with commuting and anti-commuting parameters, they constructed an extension of the Poincar\'e group which included fermionic supercharges with anticommuting parameters and the internal symmetry group, thus bypassing the Coleman-Mandula no-go theorem. 
 
The starting point for their work was the question of whether Goldstone particles with spin one half might exist, 
 in order to explain the apparent masslessness of the neutrino by interpreting it as a Goldstone fermionic particle. As Volkov recollects (2000, 58), this idea was inspired by work of Heisenberg in the second half of the 1960s, where, in the framework of a general program for a unified field theory of elementary particles, he conjectured that the neutrino could be a Goldstone particle emerging as a result of spontaneous symmetry breaking. In Volkov's words, ``This idea of Heisenberg was revolutionary, because he was the first to formulate the thought that there might exist Goldstone particles in nature with spin one-half'' ({\it ibid.}). 
 
 Heisenberg's theory was not correct but the idea motivated Volkov to consider how Goldstone fermionic particles might appear in the theory, 
 realising that this required ``an extension of the Lorenz group and an extension of the Poincar\'e group so that new operators would be present, which would correspond
to a quantum number of the neutrino'' ({\it ibid.}).\footnote{Corwin, Ne'eman, and Sternberg (1975, 588) note that the hypothesis that the neutrino's masslessness might indicate that it is a Goldstone particle was suggested earlier also by Ne'eman and had failed because of the statistics issue.} On this basis, Volkov and Akulov proceeded to formulate a Lagrangian for the interaction of the neutrino (as the Goldstone fermion) with itself and with other particles,  based on a nonlinear realization of the extended Poincar\'e algebra.\footnote{The idea of
 interpreting neutrinos as Goldstone particles of a spontaneously broken (super)symmetry was then abandoned because of phenomenological difficulties. According to Marinov (2000, 167), this was one reason why the work of Volkov and Akulov went relatively unnoticed when it came out, to be revived only after the result of Wess and Zumino.}

 \bigskip

 \subsection{West}

\medskip

\noindent $\bullet$ {\bf Two-dimensional supergauge symmetry in dual models}. In parallel to the findings in Moscow and Kharkov, a completely distinct line of development leading to supersymmetry emerged in the West.

This process originated in the context of the ``founding era'' of string theory,  starting with the 
discovery by Gabriele Veneziano of his famous ``dual'' scattering amplitude for four mesons:\footnote{``Dual'', because the amplitude also obeyed the duality principle known as
Dolen-Horn-Schmid duality (DHS duality) or dual bootstrap, besides the basic principles of the $S$-matrix theory, such as unitarity, analyticity
and crossing symmetry.} that is, the so-called {\it dual theory of strong interactions} (1968-1973),
an intense theoretical activity aimed at extending the Veneziano amplitude toward more realistic models -- from the first two models for the scattering
of {\it N} scalar particles (the generalised Veneziano model, known as the Dual
Resonance Model (DRM), and the Shapiro-Virasoro
model (SVM)) to all the subsequent endeavours to enlarge, complete and refine the dual 
theory, including its interpretation in terms of a quantum-relativistic string and the addition of fermions.\footnote{A detailed reconstruction of these developments can be found in Cappelli, Castellani, Colomo and Di Vecchia (2012).}\\

\noindent (a) {\it The Ramond (R) model}. It was precisely in connection with the fermion issue within this broader effort of model building and generalization that the Western path to supersymmetry took off. 
The starting point was the construction of the first dual model including fermions by Pierre Ramond (Ramond, 1971). In his seminal paper, Ramond proposed 
a wave equation for free fermions in the framework of the operator-formalism approach to dual theory.\footnote{According to this approach, studying the spectrum of the states amounted to rewrite the {\it N}-point amplitudes as quantum-mechanical expectation values of creation-annihilation operators of the harmonic oscillator (see for example  Cappelli {\it et al.} 2012, Chapter 10).} As emphasized by Ramond, his equation was a generalization of Dirac's
equation to the Veneziano model, according to  ``a correspondence principle'' by which basic notions of point particles (such as the momentum operator and Dirac's gamma matrices) were extended in order to be related to dual models (e.g., Ramond  2012, 29.4). Moreover,  as a consequence of his construction, Ramond obtained  a ``curious algebra, which had both commutators and anticommutators'' (Ramond 2000, 5): in fact, a generalization of the so-called Virasoro algebra to a Lie algebra with both odd and even elements -- a superalgebra, in later terminology.\footnote{In constructing the spectrum of the physical states, the problem was to decouple unphysical negative-norm states (called ``ghosts'', at the time) related to the time-like components of the oscillators. This was done, in analogy with the ``Fermi condition'' in QED following from gauge invariance, by imposing ``gauge'' constraints on the physical states. Virasoro showed that the consistency of this approach resulted from the structure of an infinite-dimensional Lie algebra (for details, see for example Di Vecchia, 2012, Chap. 11).}\\

\noindent (b) {\it The Neveu-Schwarz (NS) model}. Soon after Ramond's result, Andr\'e Neveu and John Schwarz presented a new dual model for pions by 
enlarging the Fock space of the dual resonance model (DRM) with the addition of 
anticommuting creation and annihilation operators (Neveu and Schwarz, 1971). They also obtained to extend the Virasoro algebra with a set of anticommuting generators -- half odd integer labelled, in their case (while integer labelled in the R model). Shortly thereafter, Neveu, Schwarz and Thorn (1971) reformulated the dual pion model in another Fock space, which had the advantage of containing fewer spurious states, thus allowing the spectrum of physical particle states to be studied more easily. Finally,  a few days later Thorn (1971) presented a dual model for pions and fermions, obtaining for the mesons the same spectrum as the NS model, and for the fermions the spectrum of the Ramond's propagator. Thus, the two dual models  (R and  NS) could be identified as the fermionic and bosonic sectors of a single model, later known as the Ramond-Neveu-Schwarz (RNS) model.\\

\noindent (c) {\it Supergauges in dual models}. The next step towards supersymmetry was the reformulation of 
the RNS model in terms of the world-sheet approach to dual string theory by Jean-Loup Gervais and Bunji Sakita (1971).\footnote{The idea that the dual model oscillators represented a relativistic string allowed to reformulate it as a field theory in a two-dimensional space: the coordinate of the string moving in spacetime could also be
considered a field that takes values on the world-sheet of the string (the surface swept out by the string moving in spacetime).}  As  Gervais (2000, 19) recollects, ``in 1970-71 we  started to develop the world-sheet interpretation of the spinning string (unknown at that early time) ... We recognized that the Neveu-Schwarz-Ramond (NSR)
models included world-sheet two-dimensional Dirac spinor fields, in addition to the world-sheet scalar fields.''   In particular, guided by idea that the conformal invariance of the free world-sheet action was at the origin of the elimination of negative norm states (ghosts), they succeeded in interpreting the additional operators used as the subsidiary gauge conditions on the physical states of the generalized dual models -- the set of anticommuting generators extending the Virasoro algebra (the ``supergauges'' of the NSR models, in their terminology) -- as generators of infinitesimal transformations of the fields leaving the action invariant, that is the generators of a new symmetry, the super conformal invariance of the world-sheet. 
In Gervais's words, ``We showed that the supergauges of the NSR models corresponded to the fact that the two-dimensional world-sheet Lagrangian
was invariant under transformations with anticommuting parameters which mixed the scalar and spinor fields. This gave the first example
of a supersymmetric local Lagrangian (albeit, two-dimensional)'' ({\it ibid.}, 22).

\bigskip

\noindent $\bullet$ {\bf Four-dimensional supersymmetry}. As said, the supergauge symmetry found by Gervais and Sakita was only two-dimensional (among other limitations). 
Generalizing from this dual model supergauge symmetry,  Julius Wess and Bruno Zumino, at the time collaborating at CERN,\footnote{Apparently, as  Sakita recollects (2000, 31), it was a seminar given by Sakita at the CERN in 1973, and a subsequent conversation with Zumino during his visit there, that had led Wess and Zumino to start their work on supersymmetric field theory.} succeeded in obtaining -- as a ``natural'' extension -- an analogous transformation group in four-dimensional spacetime in a 1973 seminal paper (published in 1974). It is worth recalling how their paper began, summarizing prior results and setting the agenda for their own work: 

\begin{quote}

{\small Supergauge transformations have been studied until now in dual models, especially
in their formulation as two-dimensional field theories. They transform scalar (in general tensor) fields into spinors and boson fields into fermion fields. This is possible because the parameters of the supergauge transformation are themselves totally anticommuting spinors. The commutator of two infinitesimal supergauge transformations is a conformal transformation in two dimensions. Invariance under supergauge transformations is closely connected to the absence of ghost
states in the two dimensional field theory.\\
It is natural to ask whether one can define supergauge transformations in four dimensional space-time. In this paper we show that this is indeed possible [...]. (Wess and Zumino, 1974a, 39)} 

\end{quote}

\noindent Indeed, in this first paper Wess and Zumino achieved to establish supergauge transformations in four dimensions, study the algebra generated by them and give a number of representations of them as transformations on fields. In a second paper appearing a few months later (Wess and Zumino, 1974b), they proceeded to construct a Lagrangian model that was invariant under supergauge transformations in the one-loop approximation, discuss the renormalization procedure
and show that the relations among masses and couplings were preserved by renormalization. Finally, with a third closely following paper (Wess and Zumino, 1974c), they addressed the issue of whether quantum electrodynamics could be extended, by adding suitable fields, to a renormalizable supergauge invariant theory. They succeeded to show that this was indeed possible, by constructing a field theory which was both supergauge invariant and invariant under ordinary gauge transformations. 

Since the results anticipating supersymmetry which had been obtained in the Soviet Union were still not known in the West, it was with 
these three papers that supersymmetry was really brought to the attention of the international community of theoretical
particle physicists.\footnote{This can be checked also in a graph to be found in comment by Shifman on the chronology and numerology of research in supersymmetry, reporting the number of papers on supersymmetry published between 1971 and
1982. For what regards the first part of the graph, the beginnings of the accelerated growth in 1973 is very visible (Shifman 2000, 43). The graph is also reprinted in Shifman (2025, 12)}

Note that, in these papers, Wess and Zumino still used the term `supergauge' instead of `supersymmetry', as is also evident from the papers' titles. In fact, the four-dimensional supersymmetry introduced in their work was not a gauge symmetry, although suggested by the supergauge symmetry of the dual models.\footnote{It was rather a classification symmetry for the spectrum of the states, such as the familiar $SU(N)$ symmetries, but acting in spacetime.}  But, as emphasized by Zumino, since its existence was suggested by the dual models,  ``the name supergauge symmetry in four dimensions seemed a natural choice''. However, he continued, ``it seems now reasonable to avoid the word gauge and adopt the expression Fermi-Bose supersymmetry, or simply supersymmetry, suggested recently by Salam and Strathdee'' (Zumino, 1974, 254). Apparently, as confirmed also by these words of Zumino, the term `supersymmetry' was used for the first time in the 1974 paper ``Super-symmetry and non-Abelian
gauges'' by Salam and Strathdee.\footnote{In the foreword to the first edition of their collective volume {\it The Supersymmetric World},  Kane and Shifman remark that the terminology oscillated between super-gauge and super-symmetry (Kane and Shifman, 2000, ix). In the second edition of the volume, Shifman also mentions a curious priority dispute raised by Zumino concerning the introduction of the term `supersymmetry' (see Shifman, 2025, 5).}

\section{From supersymmetry to supergravity} After the 1973/74 papers by Wess and Zumino, the theory of supersymmetry progressed very rapidly in various directions (renormalizability and `miraculous' cancellation of divergences, the superfield/superspace formalism, the construction of supersymmetric Yang-Mills theories, mechanisms for the spontaneous breaking of supersymmetry, ...).\footnote{There are numerous volumes that provide excellent coverage of these developments: for example,  Fayet and Ferrara (1977), Ferrara (1987), Kane and Shifman (2000; 2025), Duplij, Siegel, Bagger (2004), Shifman (2000).}

Here, we focus exclusively on a few key developments in local supersymmetry that led to the establishment of supergravity:  from the pioneering works of Volkov and Soroka  (1973-1974) in the Soviet Union,  to the two seminal papers for supergravity --  Freedman, van Nieuwenhuizen and Ferrara (1976), and Deser and Zumino (1976), respectively. 

\medskip

\noindent $\bullet$ {\it Towards supergravity}. After considering globally supersymmetric field theory in four
dimensions, it was natural to extend the theory by searching for local realizations of supersymmetry.  Aside from the natural inclination to generalize, which is inherent in the scientific process, there were two main (interconnected) motivations at play: the transfer to the case of supersymmetry of the `gauge ideology' of quantum field theory (gauge principle, spontaneous symmetry breaking and Higgs mechanism) and the suggestion that local supersymmetry could lead to a generalized (possibly finite) theory of gravity.\footnote{A suggestion based on the fact that field theories with local supersymmetry generalized Einstein's general relativity and that in supersymmetric field theories the loop contributions of fermions and bosons tend to cancel each other.} 

Hints in this direction were already present in the concluding programmatic remarks of the 1972 paper by Volkov and Akulov (cf. 2.2): ``Similarly, the gravitational interaction may be included by means of introducing the gauge fields for the Poincar\'e group. Note that if the gauge field for the fermionic transformation is also introduced,
then as a result of the Higgs effect the massive gauge field with spin three-halves appears [the graviton's superpartner] and the considered Goldstone particle with spin one-half [the Goldstino] disappears''.  This program was then effectively carried on by Volkov and another of his students, Vyacheslav Soroka, in two subsequent papers devoted to the Higgs mechanism for Goldstone particles with spin 1/2, later known as the ``super-Higgs effect''  (Volkov and Soroka, 1973; 1974). 

Shortly after the introduction of the Wess-Zumino model, the research of Volkov and his collaborators gained recognition in the West, significantly influencing subsequent developments toward supergravity.\footnote{Zumino (1974, 258), for example, referred to the 1973 work of Volkov and Soroka in the following terms:``Volkov and Soroka have developed a description of curved superspace which combines gravitational theory with the interaction of particles of spin 3/2, 1 and 1/2. Can a theory of this kind, because of the compensation of divergences due to supersymmetry, provide a renormalizable description of gravitational interactions?''.} 
In particular, with a paper on gauge fields on superspace (Volkov, Akulov and Soroka, 1975)
they also contributed to develop the so-called superspace approach to supergravity, initiated by  Arnowitt, Nath and Zumino (1975) on the grounds of seminal work on the introduction and use of the concept of superfield by Salam and Strathdee (1974a,b).\footnote{In these seminal works the notion of superfield (a field defined on an enlarged space, called ``superspace'') was introduced and the supersymmetry transformations were interpreted as operations on superfields. This allowed to develop a geometric approach to supergravity, based on a group-theoretic approach {\it via} curvatures (e.g., Freedman, 1978, 536).}  
 
\noindent $\bullet$ {\it 1976: The `discovery' of supergravity}. Two nearly simultaneous papers appeared in 1976 -- one by Daniel Freedman, Peter van Nieuwenhuizen, and Sergio Ferrara, and another by Stanley Deser and Bruno Zumino -- are widely recognized as marking the birth of supergravity: that is, the construction of the first theory of supergravity as a gauge theory of local supersymmetry. While the same theory was constructed by the requirement that a new gauge field, carrying spin 3/2 (gravitino), could be coupled to Einstein gravity with a locally supersymmetric interaction, the two papers remarkably differed in their approach.\footnote{Deser (2018) provides an interesting, first-hand comparison of the modalities, timing and approaches of the two papers.} Freedman, van Nieuwenhuizen and Ferrara used a second order formalism for gravitation and computer calculations for checking the vanishing of complicated terms in deriving the invariance of the action; Deser and Zumino used a simplified first order formulation of the theory, implying a torsion contribution to the geometry.\footnote{This formulation was closely related to the description of supergravity in superspace (see Deser and Zumino, 1976,  335.)}

 \section{Conclusion: the ``Dirac mode of quest''}
 
 It is widely held that the developments leading to supersymmetry in the early 1970s were a ``purely intellectual achievement, driven by the logic of theoretical development rather than by the pressure of existing data'',  
 to quote Gordon Kane and Mikhail Shifman in the foreword to their remarkable 2000 collective volume on the beginnings of the theory. According to them, this makes the history of supersymmetry  ``exceptional'',  setting it apart from ``all other major conceptual developments in physics and science ... occurred because scientists were trying to understand or study some established aspect of nature, or to solve some puzzle arising from data.''  
 
Here, Kane and Shifman (among many others) appear to conflate the notion that a scientific development can be driven by theoretical considerations rather than empirical pressure with the idea that it is  ``purely intellectual''.  But scientific practice encompasses both theoretical and experimental work, and these two facets do not always advance in tandem.

Certainly, this perspective on the intellectual character of supersymmetry is influenced by the persistent lack of empirical confirmation -- no superparticles have been detected thus far.  However, even in that regard, supersymmetry is not an outlier in the history of science: just think of the case of the Higgs boson (theoretically introduced in 1964, experimentally `observed' in 2012).  

How to trust a theory in absence of empirical confirmation is a key issue, much debated in current philosophy of science (see, for example, Dardashti, Dawid and Th\'ebault, 2019). Yet, this is a different question from whether a theoretical advance is scientific or merely an intellectual exercise. As outlined in the preceding sections, the developments of supersymmetry and supergravity are just as `scientific' as a great deal of theoretical physics. Their history is not an ``exception'':  it provides a clear example of how scientific progress can arise through the concurrent processes of generalization, analogy, and conjecture, supported by a tight interplay of mathematically
driven creativity and physical constraints and motivations. 

Reflecting on the directions of particle physics, Yoichiro Nambu (1985, 105)  distinguished two competing ``modes of quest''  -- the ``Yukawa mode'' and the ``Dirac mode'' -- in the efforts of theorists:

\begin{quote}

{\small The Yukawa mode is the pragmatical one of trying to divine
what underlies physical phenomena by attentively observing them, using available theoretical
concepts and tools at hand.
 
 The other mode, the Dirac mode, is to invent, so to speak, a new mathematical concept or
framework first, and then try to find its relevance in the real world [...] }

\end{quote}

\noindent Among examples of the Dirac mode, ``unique to physics among the natural sciences'', he included magnetic monopoles, non-Abelian gauge theory and supersymmetry. 
And he concluded:

\begin{quote}

{\small  On rare occasions, these two modes can become one and the same, as in the cases of the
Einstein gravity and the Dirac equation.}
 
\end{quote}

\bigskip

\bigskip

\bigskip

\begin{center}

{\bf References}

\end{center}

\medskip

Akulov, V. P., Volkov, D. V. and Soroka, V. A. (1975). Gauge Fields on superspace with different holonomy groups. {\it JETP Lett.} 22, 187-188.\\

Berezin, F. A. and  Katz, G. I. (1970). Lie groups with commuting and anti-commuting parameters. {\it Math. USSR-Sb.} 11, 311-325.\\

Cappelli, A., Castellani, E., Colomo, F. and Di Vecchia, P. (Eds.) (2012). {\it The birth of string
theory}. Cambridge: Cambridge University Press.\\

Corwin, L, Ne'eman, Y. and Sternberg, S. (1975).  Graded Lie Algebras in Mathematics and Physics
(Bose-Fermi Symmetry). {\it Rev. Mod. Phys.} 47, 3: 573-603.\\

Coleman, S. and Mandula, J. (1967). All possible symmetries of the S matrix. {\it Phys. Rev.}
159, 5: 1251-1256.\\

Dardashti, R., Dawid, R.  and Th\'ebault, K. (Eds.). (2019). {\it Why trust a theory? Epistemology
of fundamental physics}. Cambridge: Cambridge University Press.\\

Deser, S., and Zumino, B. (1976). Consistent supergravity. {\it Phys. Lett.} 62B, 335-337.\\

Deser, S. (2018). A brief history (and geography) of Supergravity: the first 3
weeks...and after. {\it Eur. Phys.} J. H 43, 281-291.\\

Di Vecchia, P. (2012). From the $S$-matrix to string theory. In In A. Cappelli, E. Castellani, F. Colomo and P. Di Vecchia, P. (Eds.), 
{\it The birth of string theory}. Cambridge: Cambridge University Press, Chap. 11.\\

Duplij, S.,  Siegel, W., and Bagger, J. (Eds) (2004). {\it Concise Encyclopedia of Supersymmetry}. Dordrecht: Kluwer.\\

Fayet, P. and Ferrara, S. (1977). Supersymmetry. {\it Phys. Reports.} 32, 249-334.\\

Ferrara, S. (Ed.) (1987). {\it Supersymmetry}, Singapore: World Scientific.\\

 Freedman, D. Z., van Nieuwenhuizen, P. and Ferrara, S. (1976). Progress toward a theory of supergravity.
{\it Phys. Rev.} B113, 3214-3218.\\

Freedman, D. Z. (1978). Review of Supersymmetry and Supergravity. {\it Phys. Soc. Japan}, 535-548.\\

Gervais, J.-L. and Sakita, B. (1971). Field theory interpretation of supergauges In dual models, {\it Nucl. Phys.} B34, 632-639.\\

Gervais, J.-L. (2000). Remembering the Early Times of Supersymmetry. In G. Kane and M. Shifman (Eds.),  {\it The Supersymmetric World: The Beginnings of the Theory}. 
Singapore: World Scientific, 19-25. \\

Gol'land, Yu. A.  and Likhtman, E. P. (1971). Extension of the Algebra of  Poincar\'e Group Generators and Violation of P Invariance. {\it JETP
Lett.} 13: 323-326\\

Golfand, Yu. A. and Likhtman, E. (2000). On $N =1$ symmetry algebra and simple models.  In M. Shifman (Ed.), {\it The Many Faces of the Superworld. Yuri Golfand Memorial Volume},      Singapore: World Scientific., 54-85.\\

Haag, R., Lopuszanski, J. and M. Sohnius, M. (1975). All Possible Generators of Supersymmetries of the S-Matrix. {\it Nucl. Phys.} B88, 257-274.\\

Kane, G. and Shifman, M. (2000) (Eds.). {\it The Supersymmetric World: The Beginnings of the Theory}. Singapore: World Scientific.\\

Likhtman, E. (2000). Notes of an old graduate student. In G. Kane and M. Shifman (Eds.),  {\it The Supersymmetric World: The Beginnings of the Theory},
Singapore: World Scientific, 34-41.\\

Marinov, M. S. (2000). Revealing the path to the superworld. In G. Kane and M. Shifman (Eds.),  {\it The Supersymmetric World: The Beginnings of the Theory},
Singapore: World Scientific, 158-168.\\

Nambu, Y. (1985). Directions of Particle Physics. {\it Progr. Theor. Phys. Suppl.} 85, 104-110.\\

Neveu, A.  and Schwarz, J.H. (1971). Factorizable Dual Model of Pions. {\it Nucl. Phys.} B31, 86-112. \\

Neveu, A., Schwarz, J.H. and Thorn, C. B. (1971). Reformulation of the Dual Pion Model. {\it Phys. Lett.} 35B, 529-533.\\

Nijenhuis, A. (1955). Jacobi-Type Identities for Bilinear Differential Concomitants of Certain Tensor
Fields. {\it Indagationes Mathematicae} 58, 390-397.\\

Ramond, P. (1971), Dual Theory for Free Fermions, {\it Phys. Rev.} D3, 2415-2418.\\

Ramond, P. (2012). Dual model with fermions: memoirs of an early string theorist. In A. Cappelli, E. Castellani, F. Colomo and P. Di Vecchia, P. (Eds.), 
{\it The birth of string theory}. Cambridge: Cambridge University Press, Chap. 29.\\

Sakita, B. (2000). Reminiscences. In G. Kane and M. Shifman (Eds.),  {\it The Supersymmetric World: The Beginnings of the Theory}. 
Singapore: World Scientific, 26-31.\\

Salam, A. and Strathdee, S. Super-symmetry and non-Abelian gauges. {Phys. Lett.} 51B, 353-355. \\

Shifman. M. (Ed.), {\it The Many Faces of the Superworld. Yuri Golfand Memorial Volume}, Singapore: World Scientific. \\

Shifman, M. (2000). Comment on the chronology and numerology of research on supersymmetry. In M. Shifman (Ed.), {\it The Many Faces of the Superworld. Yuri Golfand Memorial Volume}, Singapore: World Scientific., 41-43.\\

Shifman, M. (2025). Second edition of  G. Kane and M. Shifman (Eds.). {\it The Supersymmetric World: The Beginnings of the Theory}. Singapore: World Scientific.\\

Thorn, C. (1971). Embryonic dual model for pions and fermions. {\it Phys. Rev.} D4, 1112-1116.\\

Varadarajan, V. S. (2010). Introduction. In S. Ferrara, R. Fioresi and V. S. Varadarajan (Eds.), {\it Supersymmetry in Mathematics and Physics}.  Los Angeles: Springer.\\

Volkov, D. V., and Akulov, V. P. (1972). Possible Universal Neutrino Interaction. {\it JETP Lett.} 16, 438-440.\\

Volkov, D. V., and Akulov, V. P. (1973). Is the Neutrino a Goldstone Particle? {\it Phys. Lett.} 46B, 109-110. \\

Volkov, D. V. and Soroka, V. A. (1973). Higgs Effect for Goldstone Particles with Spin 1/2. {\it JETP Lett.} 18,  312-314.\\

Volkov, D. V. and Soroka, V. A. (1974).  Gauge fields for symmetry group with spinor parameters. {\it Theor. Math. Phys.} 20, 829-834.\\

Volkov, D. V. (2000). Lst interview with D. V. Volkov. In G. Kane and M. Shifman (Eds.),  {\it The Supersymmetric World: The Beginnings of the Theory}. 
Singapore: World Scientific, 54-62.\\

Wess, J. and Zumino, B. (1974a). Supergauge Transformations in Four Dimensions. {\it Nucl. Phys.} B70, 39-50. \\

Wess, J. and Zumino, B. (1974b). A Lagrangian Model Invariant under Supergauge Transformations. {\it Phys. Lett.} B 49, 52-54. \\

Wess, J. and Zumino, B. (1974c). Supergauge Invariant Extension of Quantum Electrodynamics. {\it Nucl. Phys.} B78, 1-13. \\

Wess, J. (2000). From Symmetry to Supersymmetry. In G. Kane and M. Shifman (Eds.),  {\it The Supersymmetric World: The Beginnings of the Theory}. 
Singapore: World Scientific, 67-86.\\

Zumino, B. (1974). Fermi-Bose Supersymmetry. Supergauge Symmetry in Four Dimensions. In J. R. Smith and G. Manning (Eds.), {\it Proc. ICHEP '74}, 
Chilton: Rutherford Appleton Laboratory.\\ 

Zumino, B. (2006). Supersymmetry Then and Now. {\it Fortschr. Phys.} 54, 199-204.

 \end{document}